\documentclass[conference]{IEEEtran}
\IEEEoverridecommandlockouts
\usepackage{cite}
\usepackage{amsmath,amssymb,amsfonts}
\usepackage{algorithmic}
\usepackage[ruled,vlined]{algorithm2e}
\usepackage{graphicx}
\usepackage{textcomp}
\usepackage{xcolor}
\def\BibTeX{{\rm B\kern-.05em{\sc i\kern-.025em b}\kern-.08em
    T\kern-.1667em\lower.7ex\hbox{E}\kern-.125emX}}

\usepackage{tabularx,tablefootnote,wrapfig,multirow}
\usepackage{hyphenat,xspace,color,colortbl,enumitem}

\usepackage{amsmath,amsfonts,amssymb}
\usepackage{mathtools}
\usepackage{commath}
\usepackage[sc,osf]{mathpazo}

\definecolor{Gray}{gray}{0.9}
\definecolor{LightGreen}{rgb}{0.88,1,0.88}
\definecolor{LightOrange}{rgb}{1,0.85,0.8}
\definecolor{LightRed}{rgb}{1,0.80,0.80}

\newcommand{\parahead}[1]{\vspace{2pt plus 1pt minus 1pt}\noindent{\bfseries #1}}

\setitemize{itemsep=1pt,topsep=2pt,parsep=1pt,partopsep=0pt,leftmargin=1.5em}
\setenumerate{itemsep=1pt,topsep=2pt,parsep=1pt,partopsep=0pt,leftmargin=1.5em}
\setlist{itemsep=1pt,parsep=1pt}

\usepackage[english]{babel}
\usepackage{graphicx,booktabs,physics,amsmath,times}
\usepackage{tabularx,tablefootnote,wrapfig,multirow}
\usepackage{hyphenat,xspace,color,colortbl,enumitem}
\usepackage[labelformat=simple,skip=0pt]{subcaption}
\newcommand\blfootnote[1]{%
  \begingroup
  \renewcommand\thefootnote{}\footnote{#1}%
  \addtocounter{footnote}{-1}%
  \endgroup
}

\usepackage{fancyhdr}
\fancypagestyle{firstpage}{%
  \rhead{Accepted article to appear in the proceedings of IEEE ICC ’21}
}

\begin{document}
\title{Quantum Annealing for Large MIMO Downlink Vector Perturbation Precoding}



\author{\IEEEauthorblockN{Srikar Kasi$^{\star}$\\ \textit{Student Member}, \textit{IEEE}\thanks{$\star$ Co-primary authors, appearing in randomly chosen order.}}
\IEEEauthorblockA{Princeton University} 
\and
\IEEEauthorblockN{Abhishek Kumar Singh$^{\star}$\\ \textit{Student Member}, \textit{IEEE}}
\IEEEauthorblockA{Princeton University} 
\and
\IEEEauthorblockN{Davide Venturelli}
\IEEEauthorblockA{Universities Space\\Research Association}
\and
\IEEEauthorblockN{Kyle Jamieson\\\textit{Senior Member}, \textit{IEEE}}
\IEEEauthorblockA{Princeton University} 
}

\maketitle
\thispagestyle{firstpage}
\blfootnote{\textcopyright 2021 IEEE. Personal use of this material is permitted. Permission from IEEE must be obtained for all other uses, in any current or future media, including reprinting/republishing this material for advertising or promotional purposes, creating new collective works, for resale or redistribution to servers or lists, or reuse of any copyrighted component of this work in other works.}

\begin{abstract}

In a multi-user system with multiple antennas at the base station, precoding techniques in the downlink broadcast channel allow users to detect their respective data in a non-cooperative manner. Vector Perturbation Precoding (VPP) is a non-linear variant of transmit-side channel inversion that perturbs user data to achieve full diversity order. While promising, finding an optimal perturbation in VPP is known to be an NP-hard problem, demanding heavy computational support at the base station and limiting the feasibility of the approach to small MIMO systems. This work proposes a radically different processing architecture for the downlink VPP problem, one based on Quantum Annealing (QA), to enable the applicability of VPP to large MIMO systems. Our design reduces VPP to a quadratic polynomial form amenable to QA, then refines the problem coefficients to mitigate the adverse effects of QA hardware noise. We evaluate our proposed QA based VPP (QAVP) technique on a real Quantum Annealing device over a variety of design and machine parameter settings. With existing hardware, QAVP can achieve a BER of $10^{-4}$ with 100$\mu$s compute time, for a 6$\times$6 MIMO system using 64 QAM modulation at 32 dB SNR.

\end{abstract}

\begin{IEEEkeywords}
Vector Perturbation, Downlink Precoding, Quantum Computation, Quantum Annealing, Optimization
\end{IEEEkeywords}

\section{Introduction}
\label{s:intro}

Modern wireless networks are experiencing tremendous growth 
in traffic loads at base stations, and hence to meet the resulting 
computational and latency requirements, 
designers continue to investigate new architectures 
and hardware for today's 5G and 
tomorrow's 6G networks. A large component of cellular baseband processing comprises of downlink data traffic due to a significant rise in the popularity and usage of video streaming platforms (\textit{e.g.,} Netflix). To meet the ever-growing user demand, it is critical for the base stations to enhance the quality of downlink data streams in terms of throughput, error rate, and latency.


In a multi-user multiple-input multiple-output (MIMO) downlink data transmission, \textit{precoding} techniques can be used to eliminate the effect of inter-user interference and allow users to detect their respective data non-cooperatively, minimizing error-rate and maximizing throughput. In this work, we focus on \textit{Vector Perturbation Precoding} \textit{(VPP)} \cite{peel2005vector}. VPP is a widely studied non-linear precoding technique that performs transmit-side channel inversion over a perturbed user data vector to reduce the transmit power scaling. Although VPP has been shown to achieve better error performance compared to other precoding techniques (\textit{e.g.,} zero-forcing, Tomlinson-Harashima Precoding~\cite{thp}), finding an optimal perturbation for user data in VPP is known to be NP-hard, making its implementation in massive/large MIMO systems to be infeasible.

A promising and cost-effective architecture to address 
the increased computational burden of wireless networks
is a \textit{Centralized Radio Access Network 
(C-RAN)} \cite{checko2014cloud} architecture, 
which aggregates the computationally\hyp{}demanding
processing at many wireless base stations. Since baseband physical\hyp{}layer processing is highly
time\hyp{}critical and the required stream of baseband signal samples
has a high data\hyp{}rate, this type of C-RAN deployment imposes both
latency and bandwidth requirements on the interconnect between each
base station and the data center. Related studies to this end are investigating quantum computation for solving networking problems in the uplink: such as Channel coding\cite{10.1145/3372224.3419207} and ML detection\cite{qaMIMO1}. However, investigating quantum computation for problems in the downlink remains unexplored to the best of our knowledge.

In this paper, we propose a radically different processing to the NP-hard VPP problem in the downlink: \textit{Quantum Annealing based Vector Perturbation Precoding} \textit{(QAVP)}. Our proposed technique leverages recent advances in quantum computational devices and applies them to the problem of VPP. QAVP's approach is to represent the VPP problem as an optimization problem over a quadratic polynomial with binary variables \textit{i.e.}, \textit{Quadratic Unconstrained Binary Optimization} (QUBO),
which is an optimization form that a \textit{Quantum Annealer (QA)} machine takes
as input, then refine the QUBO
coefficients to mitigate the adverse 
effects of QA hardware noise and the process of mapping the polynomial onto the physical QA qubit hardware topology.  After these \emph{preprocessing} and \emph{embedding} steps,
QAVP uses a real Quantum Annealing machine to solve the resultant QUBO problem and then constructs the VPP perturbation from the solutions returned by the QA machine. Our results show that, from the standpoint of computation time, QAVP can outperform popular encoding algorithms for large MIMO systems. 






\section{Vector Perturbation Precoding}
\label{sec:downlink}

We consider a general MIMO downlink scenario where a base station equipped with $N_t$ transmit antennas communicate data streams with $N_r$ ($\leq$~$N_t$) single-antenna non-cooperative users independently and simultaneously.

In VPP, the user data symbol vector \textbf{u} $\in$ $\mathbb{C}^{N_r\times 1}$ is perturbed by an integer vector \textbf{v} $\in$ $\mathbb{G}^{N_r\times 1}$. This maps the data symbols to a wider constellation space, forming a perturbed transmit vector, \textbf{d} = \textbf{u} + $\tau$\textbf{v}. Here, \textbf{v} is a vector of Gaussian integers\footnote{The set of Gaussian integers $\mathbb{G}$ = $\mathbb{Z}$ + j$\mathbb{Z}$ consists of all complex numbers whose real and imaginary parts are integers.} and $\tau$ = 2( $\abs{c_{max}}$ + $\Delta$/2) is a constant chosen to provide symmetric decoding regions around the constellation points. $\abs{c_{max}}$ is the magnitude of the largest constellation symbol and $\Delta$ is the spacing between the constellation symbols \cite{vpPaper}. The perturbed vector \textbf{d} is then precoded with a precoder matrix (\textbf{P} $\in$ $\mathbb{C}^{N_t\times N_r}$), where the choice of \textbf{P} inverts the wireless channel \textbf{H}$\in$ $\mathbb{C}^{N_r\times N_t}$ and reduces the effect of wide range of eigenvalues of the channel coefficients \cite{vpPaper}. The precoder matrix \textbf{P} = $\textbf{H}^H(\textbf{H}\textbf{H}^H)^{-1}$ is used in Zero Forcing (ZF) precoding. The received symbol vector \textbf{y} corresponding to the transmitted symbol vector \textbf{x} = $\textbf{P}\textbf{d}/\sqrt{P_t}$  is given by,
\begin{equation}
\textbf{y} = \frac{1}{\sqrt{P_t}}(\textbf{H}\textbf{P}\textbf{d}) + \textbf{n}
\label{eq:received}
\end{equation}

where the scalar $P_t$ = $\norm{\textbf{P} \textbf{d}}^2$ is the transmission power scaling factor, which is assumed to be known at the receiver, and \textbf{n} is wireless channel noise. The receiver decodes by applying a modulo $\tau$ operation to the received signal \textbf{y}. The transmission power scaling ($P_t$) causes noise amplification during the decoding process. In order to minimize the transmission power scaling, an optimal choice of the perturbation vector ($\textbf{v}^\star$) that leads to the smallest $P_t$ is computed as in \cite{vpPaper}:
\begin{equation}
\textbf{v}^\star = \text{arg} \min_{\textbf{v}} \norm{\textbf{H}^H(\textbf{H}\textbf{H}^H)^{-1}(\textbf{u} + \tau \textbf{v})}^2.
\label{eq:vstar}
\end{equation}

\section{Related Work}
\label{sec:relatedWork}

The Sphere encoder \cite{vpPaper} builds on the Fincke and Pohst algorithm \cite{finPohst} by expressing the precoding matrix $\mathbf{P}$ as $\mathbf{P} = \mathbf{Q}\mathbf{R}$ by QR decomposition, where $\mathbf{Q}$ is unitary and $\mathbf{R}$ is upper triangular. It then performs a tree search, utilizing the upper triangular structure and limiting the search to the points within a hyper-sphere of a suitably chosen radius, and hence avoids exhaustive search over all possibilities. This algorithm is very similar to the Sphere decoder used for the MIMO receiver. While Sphere encoder is much better than exhaustive search, its expected complexity is still exponential. Park~\textit{et~al.}~\cite{decopVP} provide an approximation to the VPP problem by minimizing the real and imaginary parts of the VPP cost function separately. This reduces the complexity of VPP computation at the cost of error performance. The Thresholded Sphere Search algorithm \cite{vppThreshOpt} imposes an additional stopping criteria to Sphere encoder algorithm~\cite{vpPaper} based on an SNR dependent threshold heuristic. Another scheme for reducing the search space of Sphere encoder is discussed in \cite{modLoss}. It restricts the values of each perturbation to four possible values (two for real and two for imaginary) and hence reducing the search space of the Sphere encoder. Despite improvements in computation costs over Sphere encoder, the search space of all these methods is still exponential with respect to the MIMO size.

Several approximations for VPP (with polynomial complexity) exist in literature. The Fixed Complexity Sphere Encoder (FSE), adapts the Sphere encoder to have a lower and fixed polynomial complexity by pruning a large number of branches during the tree search but leads to degradation in error performance \cite{vpFSE}. Degree-2 Sparse Vector Perturbation (D2VP), presented in~\cite{sparseVP}, is a low complexity algorithm for vector perturbation precoding. It reduces the complexity of finding the VPP solution by assuming that only 2 elements of the perturbation vector can be non-zero and then improves the solution over multiple iterations.

\section{Quantum Annealing}
\label{sec:qa}

Quantum Annealing (QA) is a heuristic approach that implements in hardware a quantum computing algorithm inspired by the Adiabatic Theorem of quantum mechanics~\cite{albash2018adiabatic}. The method aims to find the lowest energy \textit{spin configuration} (solution) of the class of \textit{quadratic unconstrained binary optimization} \textit{(QUBO)} problems in their equivalent \textit{Ising} specification. The Ising form is described by:
\begin{equation}
E = \sum_{i}h_{i}s_{i} + \sum_{i<j}J_{ij}s_{i}s_{j}
\label{eqn:qubo}
\end{equation}

where \textit{E} is the \textit{energy}, $s_i$ is a solution variable, $h_{i}$ and $J_{ij}$ are problem parameters called \textit{bias} and \textit{coupler strength} respectively. QUBO form is obtained from Eq.\ref{eqn:qubo} by a simple variable transformation ($s_i \longrightarrow 2q_i - 1$), where $s_i$ and $q_i$ represent Ising and QUBO form variables respectively. Ising form solution variables ($s_i$) take on values in $\{-1, +1\}$, and QUBO form solution variables ($q_i$) take on values in $\{0, 1\}$. The equivalent QUBO form of Eq.~\ref{eqn:qubo} is described by:
\begin{equation}
E = \sum_{i}f_{i}q_{i} + \sum_{i<j}g_{ij}q_{i}q_{j}
\label{eqn:qubo_bits}
\end{equation}


\subsection{Primer: Quantum Annealers.} 
\begin{figure}
    \centering
    \includegraphics[width=0.65\linewidth]{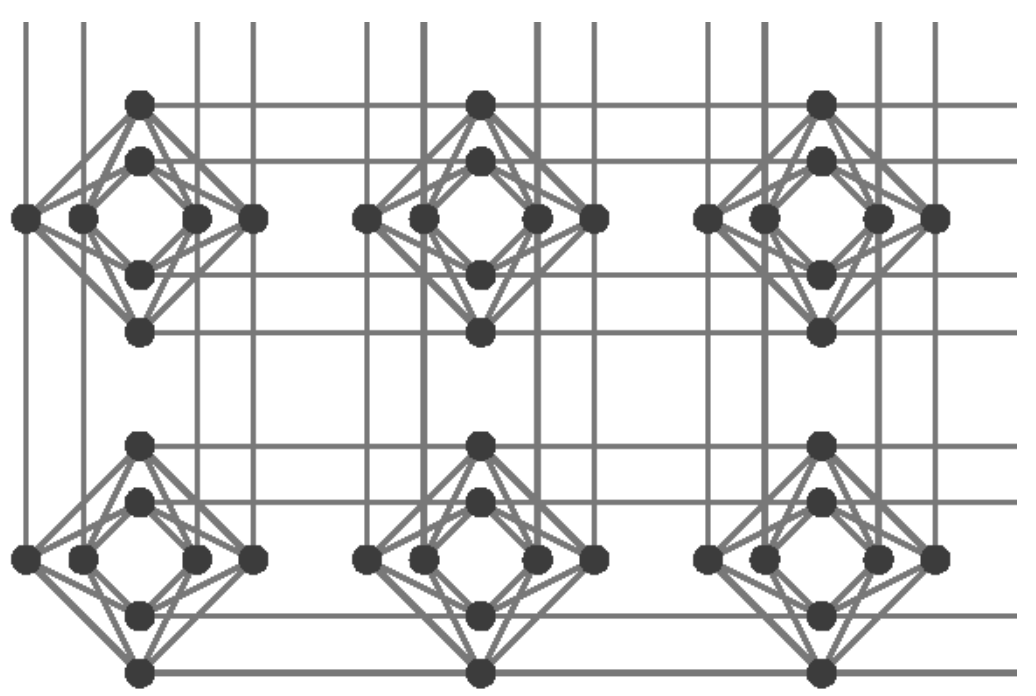}
    \caption{The figure shows \textit{qubits} (nodes) and \textit{couplers} (edges) in DW2Q QA hardware. Each group of eight qubits is called \emph{unit cell}.} 
    \label{f:chimera}
\end{figure}

Quantum Annealers are specialized quantum computers, essentially comprised of two types of resources: \textit{qubits} (quantum bits) and \textit{couplers}, whose regular connectivity structure is organized in \textit{unit cells}. Fig.~\ref{f:chimera} shows the hardware structure of the QA we adapt in this study, the D-Wave 2000Q (DW2Q) quantum annealing machine. The QA device programs the linear (\textit{i.e.,} biases) and quadratic (\textit{i.e.,} coupler strengths) coefficients of Eq.\ref{eqn:qubo} onto qubits and couplers using controllable inductive elements in proximity of superconducting Josephson Junctions present on the chip~\cite{10.1145/3372224.3419207}.

\textit{Embedding.} The process of mapping the problem at hand onto the physical QA hardware topology follows the graph-theory problem of graph minor \textit{embedding}. We demonstrate here the embedding process with an example Ising problem:
\begin{equation}
E = J_{12}s_1s_2 + J_{13}s_1s_3 + J_{23}s_2s_3
\label{eqn:example}
\end{equation}

\begin{figure}
\centering
\includegraphics[width=\linewidth]{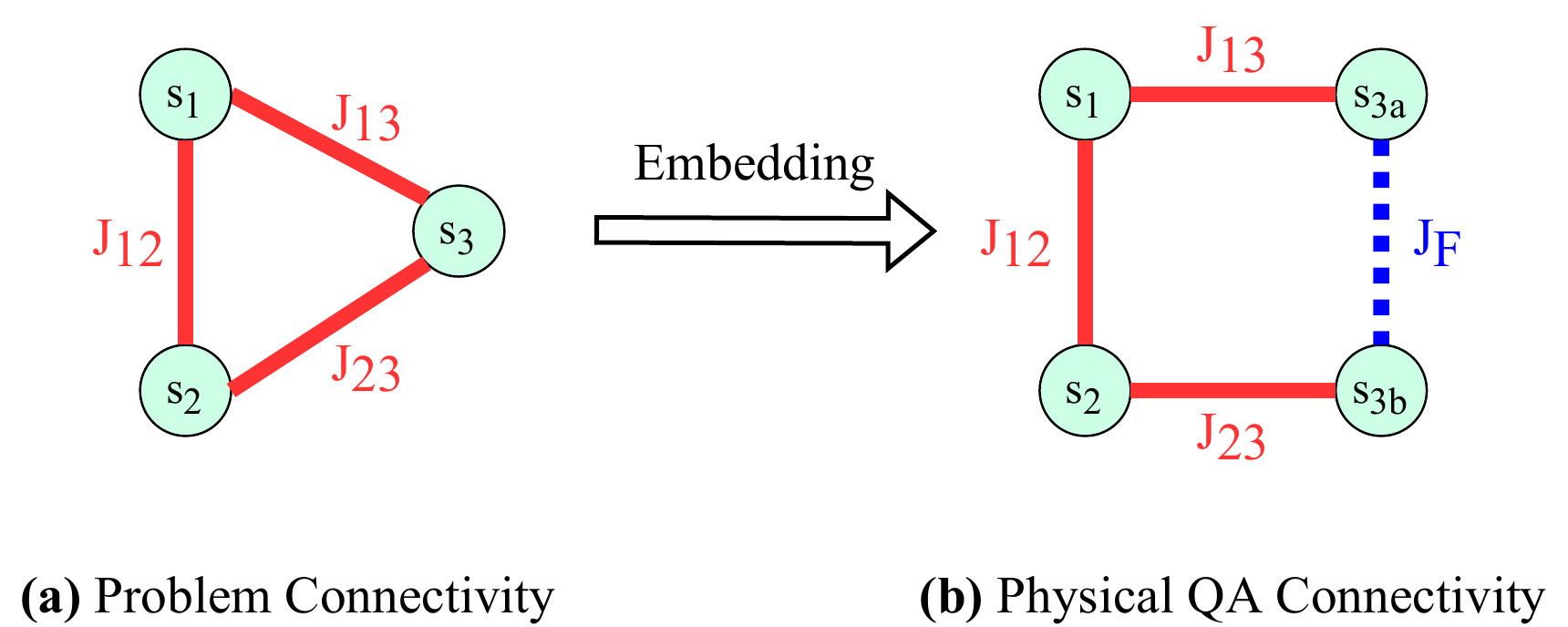}
\caption{The figure demonstrates the embedding process of Eq.~\ref{eqn:example}. In the figure, \textbf{(a)} shows the direct problem connectivity of Eq.~\ref{eqn:example} and \textbf{(b)} shows its physical connectivity on QA hardware.}
\label{fig:example}
\end{figure}

The direct connectivity of this example problem is shown in Fig.\ref{fig:example}\textbf{(a)}. However a three-node complete connectivity graph structure does not exist in the QA hardware (\textit{cf.} Fig.\ref{f:chimera}). Hence the minor-embedding approach is to map one of the problem variables in Fig.\ref{fig:example}\textbf{(a)} (\textit{e.g.,} $s_3$) onto two physical qubits (\textit{e.g.,} $s_{3a}$ and $s_{3b}$) as Fig.\ref{fig:example}\textbf{(b)} shows, such that the resulting connectivity can be realized on the QA hardware. The couplers between the qubits $s_{3a}$ and $s_{3b}$ are tasked to enforce these physical qubits to be correlated in order to end up with the same value at the end of the annealing process. This is implemented through a strong ferromagnetic interaction energy called \textit{chain strength} $J_F$  (see in Fig.\ref{fig:example}\textbf{(b)}).


\textit{Coefficient considerations.} Today's QA devices provide support for bias values in $[-2, +2]$ and coupler strength values in $[-2, +1]$ with a bit-precision guaranteed to 4--5 bits only. Further, the QA introduces an analog machine noise distinct from communication channel noise called \textit{intrinsic control errors} or \textit{ICE}. ICE noise, a collection of errors caused by qubit flux noise, susceptibility, among others \cite{dwave_tech}, essentially alters the problem coefficients ($h_{i}\rightarrow h_{i}\pm \delta h_{i}$, $J_{ij}\rightarrow J_{ij}\pm \delta J_{ij}$). Although the errors $\delta h_{i}$ and $\delta J_{ij}$ are currently on the order of $10^{-2}$, these may degrade the solution quality of some problems in scenarios where ICE noise erases significant information from the ground state of the input problem.

\subsection{Annealing process}
QA processors simulate systems in the transverse field Ising model described by the time-dependent energy functional (Hamiltonian):
\begin{equation}
H = -A(s)\sum_i\sigma_i^x + B(s)\big\{\sum_i h_{i}\sigma_i^z + \sum_{i<j}J_{ij}\sigma_i^z\sigma_j^z\big\}
\label{eqn: hamiltonian}
\end{equation}

where $\sigma_i^{x,z}$ are spin operators (Pauli matrices) acting on the $i^{th}$ qubit, $h_i$ and $J_{ij}$ are problem parameters, \textit{s} (= $t/t_a$) is called \textit{annealing schedule} where \textit{t} is the time and $t_a$ is the \textit{annealing time}. $A(s)$ and $B(s)$ are two monotonic scaling signals in the annealer such that at time $t = 0$, $A(0) \gg B(0)\approx 0$ and at time $t = t_a$, $B(1) \gg A(1)\approx 0$. The annealing algorithm initializes the system in the ground state of $\sum_i\sigma_i^x$ where each qubit is in a \textit{superposition} state $\frac{1}{\sqrt{2}} \left(\ket{0} + \ket{1}\right)$, then adiabatically evolves this Hamiltonian from time $t = 0$ until $t = t_a$ (\textit{i.e.,} decreasing \textit{A(s)} , increasing \textit{B(s)}). The time-dependent evolution described by the Schroedinger Equation driven by these signals \textit{A} and \textit{B} is essentially the annealing algorithm. While in real processor the dynamics is not ideal but it is dominated by dissipative noise instead~\cite{marshall2019power}, the expectations of the model still hold for the most part. The process of optimizing a problem in the QA is called an \textit{anneal}, while the time taken for an anneal is called \textit{annealing time}.
\section{Design}
\label{sec:design}

Quantum Annealing based Vector Perturbation (QAVP) envisions a scenario where a Quantum Annealer (QA) machine is co-located with a centralized data center for computational processing in the C-RAN architecture~\cite{checko2014cloud} which allows for low latency communication between base stations and the QA machine. QAVP converts a VPP problem to a Quadratic Unconstrained Binary Optimization (QUBO) problem, then fine-tunes the QUBO problem coefficients to mitigate the adverse effects of QA hardware noise and increase the probability of finding the correct solution with QA. The fine-tuned QUBO is next embedded onto the physical QA hardware (see Fig.\ref{f:chimera}) for running the problem. We conduct multiple anneals for a given QUBO problem, where each anneal generates a candidate solution bit-string for VPP. The QUBO solutions are converted to perturbation vectors by inverting the transform described in Eq.~\ref{eq:vpqubo}. Among these returned solutions, the solution that minimizes the VPP objective function is chosen by the base station. In scenarios where QAVP fails to find a solution with lower transmit power scaling than Zero-Forcing (ZF) precoding, the base station discards the QAVP solution and uses ZF precoding instead of VPP. 

\begin{figure}
    \centering
    \includegraphics[width=0.8\linewidth]{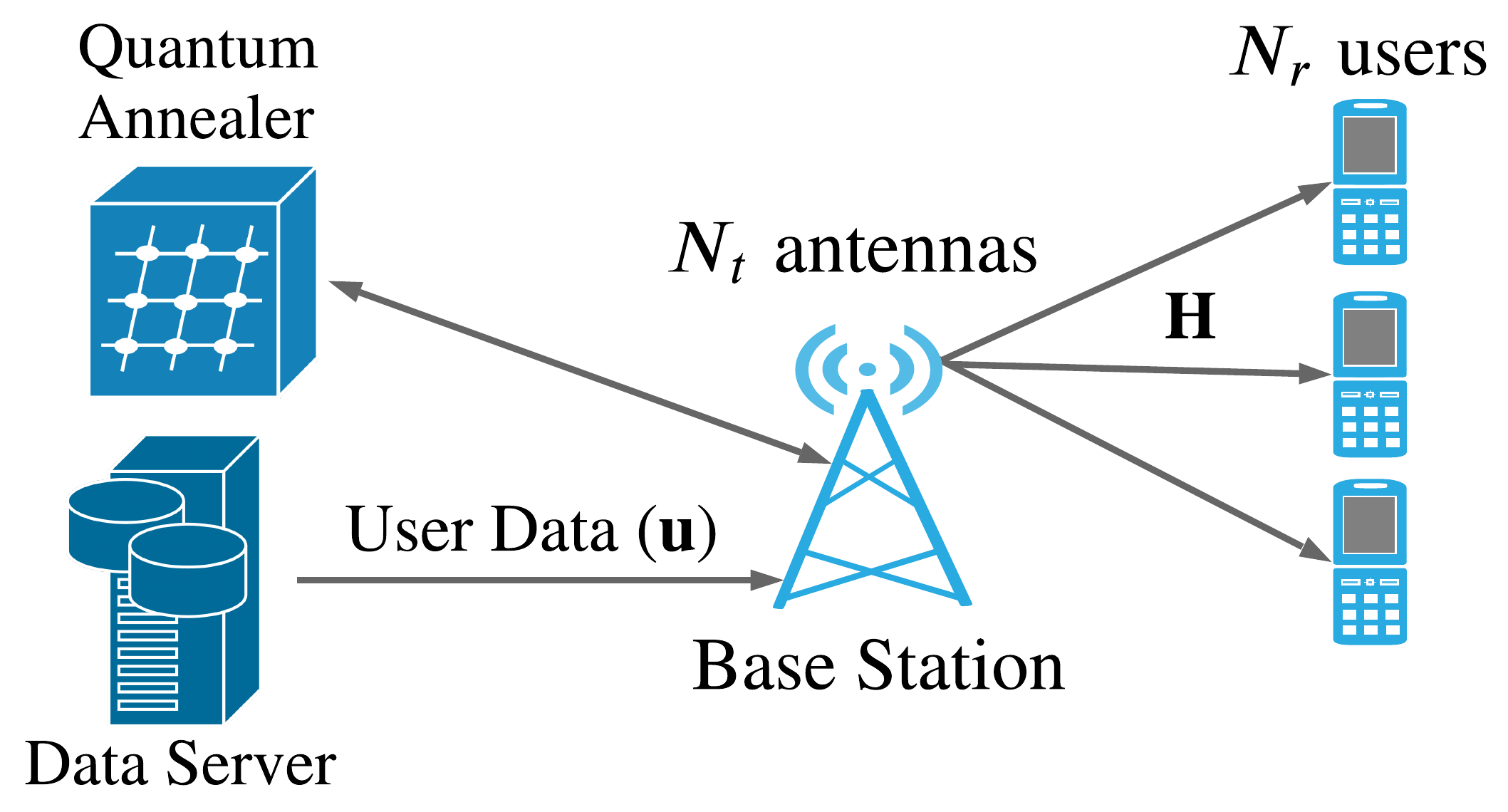}
    \caption{A typical deployment scenario of QAVP. $N_r$ users are receiving downlink data streams from a base station with $N_t$ antennas.}
    \label{fig:qavpdep}
\end{figure}

The steps involved in QAVP computation, excluding embedding and annealing, can be executed by the physical layer of base station or can be offloaded to a server in close proximity to QA machine. Fig.~\ref{fig:qavpdep} illustrates a typical deployment scenario for QAVP, where $N_r$ users are receiving downlink data streams from a base station with $N_{t}$ antennas. We next demonstrate QAVP's QUBO formulation for the VPP problem, QAVP's pre-processing, and embedding considerations.

\subsection{QAVP's QUBO Formulation} 
\label{sec:vpQubo}
The downlink VPP problem is to find an optimal perturbation vector $\textbf{v}^{\star}$ that minimizes the transmit power at the base station, whose NP-hard objective function is represented in Eq.~\ref{eq:vstar}. As the search objective in Eq.\ref{eq:vstar} is over the vector \textbf{v} $\in$ $\mathbb{G}^{N_r\times 1}$, we construct each entry in \textbf{v} by a linear combination of binary variables as follows.

Let $q_i$ denote the $i^{th}$ QUBO form solution variable, and let $a_k$ + j$b_k$ be the $k^{th}$ entry in \textbf{v}, where $a_k$ and $b_k$ are integers. We design each $a_k$ and $b_k$ $\forall k$ in an identical fashion using $t+1$ distinct solution variables as:
\begin{align}
a_k\big/b_k =  \sum_{m=1}^{t} {2^{m-1}\cdot q_m} - 2^{t}*q_{t+1}
\label{eq:vpqubo}
\end{align}

Since all the solution variables ($q_i$) are binary, this formulation allows $a_k$ and $b_k$ to take all integer values in the range $[-2^{t}, 2^{t}-1]$, where each integer corresponds to a unique configuration of solution variables. The value of \textit{t} determines the search range of \textbf{v}. Yuen~\textit{et~al.}~\cite{vpRange} show that perturbation values are most likely to be $\{-1,0,1\}$, and hence $t$ = 1 which allows perturbation values in the range $[-2,1]$, is usually sufficient. We substitute the entries of $\mathbf{v}$ with their corresponding solution variables into Eq.\ref{eq:vstar} to obtain the QAVP's QUBO:
\begin{align}
\arg\min_{\forall q} \big\{\sum_{\forall i} f_i(\textbf{H}, \textbf{u}) q_i + \sum_{\forall i, j} g_{ij}(\textbf{H}, \textbf{u}) q_iq_j\big\}
\label{eq:fqubo}
\end{align}
where $f$ and $g$ are functions of MIMO channel matrix and user data vector. The linear and quadratic coefficients obtained from Eq.~\ref{eq:fqubo} can be programmed into the QA processor. The computational complexities of the proposed QA technique and the optimal case Sphere Encoder are $O(e^{\sqrt{N_u}})$ \cite{mukherjee2015multivariable} and $O(e^{N_u})$ \cite{mohaisen2009fixed} respectively, where $N_u$ is the number of users. We next present our QUBO pre-processing considerations.

\subsection{QAVP's Pre-processing}
Our pre-processing scheme aims to mitigate the adverse effects of QA hardware ICE noise (\S\ref{sec:qa}) by scaling and eliminating minor QUBO coefficients. We note that a generic QUBO form of Eq.\ref{eqn:qubo_bits} can be equivalently expressed as:
\begin{equation}
    E = \sum_{i}f_{i}q_{i} + \sum_{i<j}g_{ij}q_{i}q_{j} = \mathbf{q^{T}}\mathbf{Q}\mathbf{q}
\end{equation}
where $\mathbf{q} = [q_{1}, q_{2}, ...]^{T}$, and $\mathbf{Q}$ is an upper triangular matrix with $Q_{ij}$ = $f_{i}$ ($i=j$), $Q_{ij}$ = $g_{ij}$ ($i<j$), $Q_{ij}$ = $0$ ($i>j$). Let $Q_{max}$ = $\max_{i,j} |Q_{ij}|$ be the maximum QUBO coefficient value, $T_{high}$ and $T_{low}$ be our chosen upper and lower bounds for the QUBO coefficient values. We define a \textit{scaleFactor} as:
\begin{equation}
scaleFactor =
  \left\{ {\begin{array}{cc}
   1 & Q_{max} \leq T_{high} \\
   \dfrac{T_{high}}{Q_{max}} & Q_{max} > T_{high} \\
  \end{array} } \right.,
\end{equation}\\
Our approach is to first scale each entry in $\mathbf{Q}$ with the scaleFactor, then eliminate coefficients below a chosen threshold of $10^{T_{low}}$. This scaling process is summarized below.

\parahead{Step 1:} Set $\mathbf{Q} \leftarrow \mathbf{Q}*scaleFactor$.\\
\parahead{Step 2:} For every $i\leq j$, if $Q_{ij}< 10^{T_{low}}$ then set $Q_{ij} \leftarrow 0$.\\ 
\begin{figure}
    \centering
    \includegraphics[width=\linewidth]{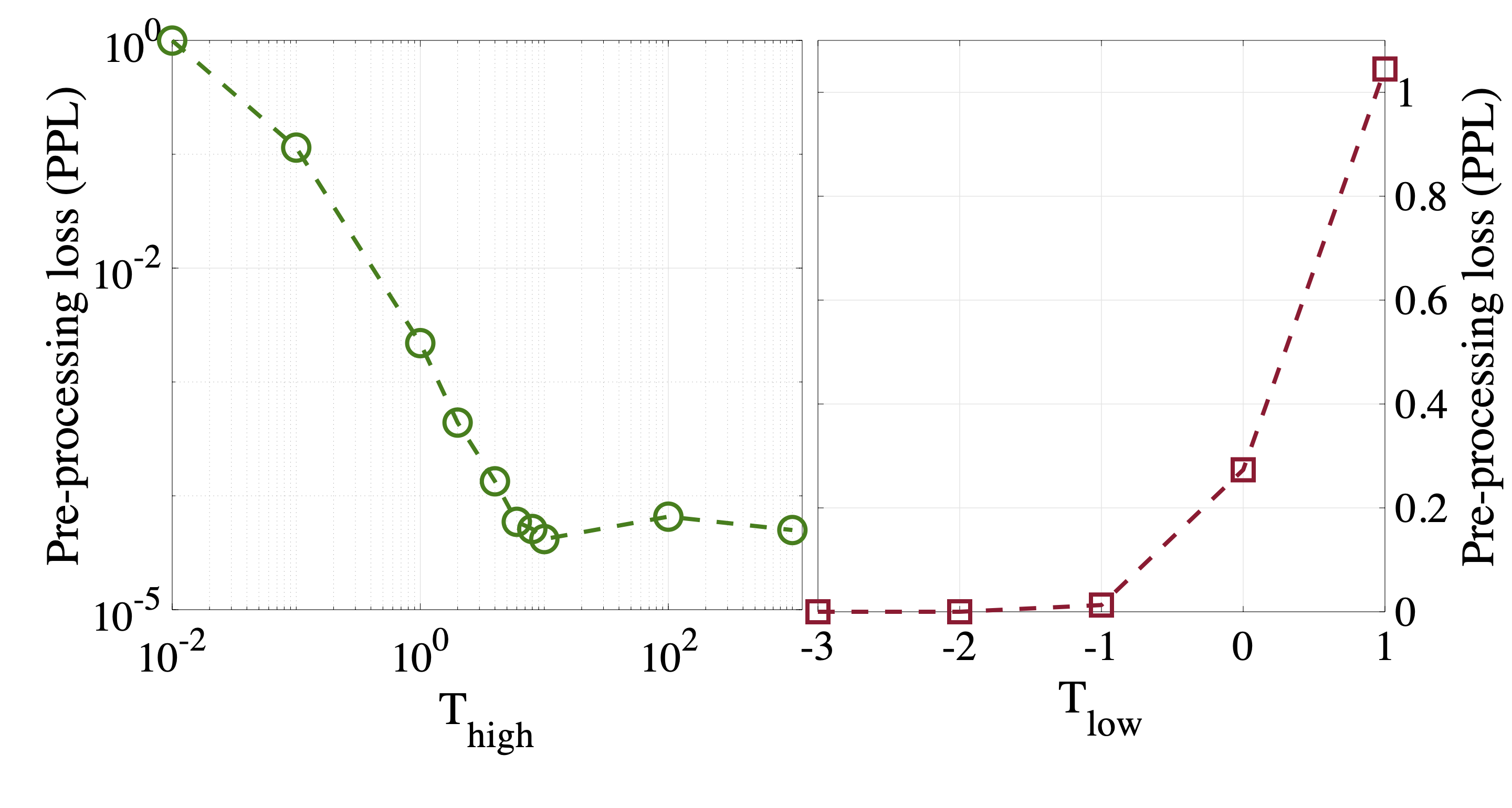}
    \caption{QAVP's pre-processing loss for different choices of $T_{high}$ and $T_{low}$, showing that $T_{high} = 6$ is near-optimal and $T_{low} = -2$ leads to negligible PPL.}
    \label{fig:quboPPL}
\end{figure}

Let $\mathbf{Q}_{pre}$ correspond to the QUBO obtained after this pre-processing step. Let $\mathbf{q}_{pre}^{\star}$ = $\arg\min_\textbf{q} \mathbf{q^{T}}\mathbf{Q}_{pre}\mathbf{q}$, and $\mathbf{q}^{\star} = \arg\min_q \mathbf{q^{T}}\mathbf{Q}\mathbf{q}$. We define a \textit{pre-processing loss} \textit{(PPL)} as:
\begin{equation}
    \text{PPL} = \dfrac{|(\mathbf{q}_{pre}^{*^T}\mathbf{Q}\mathbf{q}_{pre}^{*} - \mathbf{q^{*^T}}\mathbf{Q}\mathbf{q^{*}})|}{|(\mathbf{q^{*^T}}\mathbf{Q}\mathbf{q^{*}})|}
\end{equation}

PPL is used to quantify deterioration in the optimal value of the VPP problem due to the modification of the original QUBO problem. We see from Fig.~\ref{fig:quboPPL} that PPL increases with an increase in $T_{low}$ as a higher $T_{low}$ causes more coefficients to be reduced to zero. We further observe that PPL reduces initially with $T_{high}$, and then becomes constant. While having a high negative value for $T_{low}$ leads to low PPL, it makes it more susceptible to QA ICE noise. The optimal choices for $T_{high}$ and $T_{low}$ depend on the distribution of QUBO coefficients, magnitude of QA ICE noise, and problem embedding. Our empirical studies find that a $T_{high} = 6$ and a $T_{low} = -2$ obtains good solutions with the DW2Q QA.

\subsection{QAVP's Embedding and Parallelism.} We use a D-Wave's heuristic algorithm described in Ref.~\cite{cai2014practical} and implemented in D-Wave MinorMiner software libraries for mapping of QAVP's QUBO design onto the QA hardware. It is to note that multiple QUBOs or multiple instances of a QUBO can be parallelly processed by mapping different problems onto distinct physical locations in the QA chip. For instance, QAVP for 7$\times$7 MIMO system requires approximately 200 qubits, implying that 9--10 such problems can be simulataneously solved on the DW2Q QA with 2,048 qubits. The number of qubits in the QA hardware has been steadily doubling each year and this trend is expected to continue \cite{10.1145/3372224.3419207}, with the latest 5,000 qubit chip just released at time of writing. While the evolution of QA technology is currently at an early stage (2011-), future QA processors with higher qubit counts will potentially increase the supported parallelism for QAVP.



\section{Evaluation}
\label{sec:eval}

In this section, we first present microbenchmarks for QA machine parameters. We then investigate QAVP's end-to-end error performance, comparing head-to-head against Zero Forcing (ZF) and Fixed Complexity Sphere Encoder (FSE) algorithms. The DW2Q QA system performance is majorly affected by the choice of \textit{annealing time} ($T_a$), number of anneals ($N_a$), and \textit{chain strengths} ($J_F$). Recall that number of anneals~($N_a$) refers to the number of times annealing process is repeated for each QUBO problem. Our evaluation methodology is as follows: we obtain the VPP solutions from the QA (DW2Q), and use these solutions to simulate an end-to-end downlink data transmission on a trace-driven MIMO simulator (implemented in MATLAB). We simulate data transmission over Rayleigh fading wireless channels and empirically measure the BER corresponding to different users. 

\subsection{Effect of Chain Strength ($J_F$)}
\begin{figure}
\centering
\includegraphics[width=0.65\linewidth]{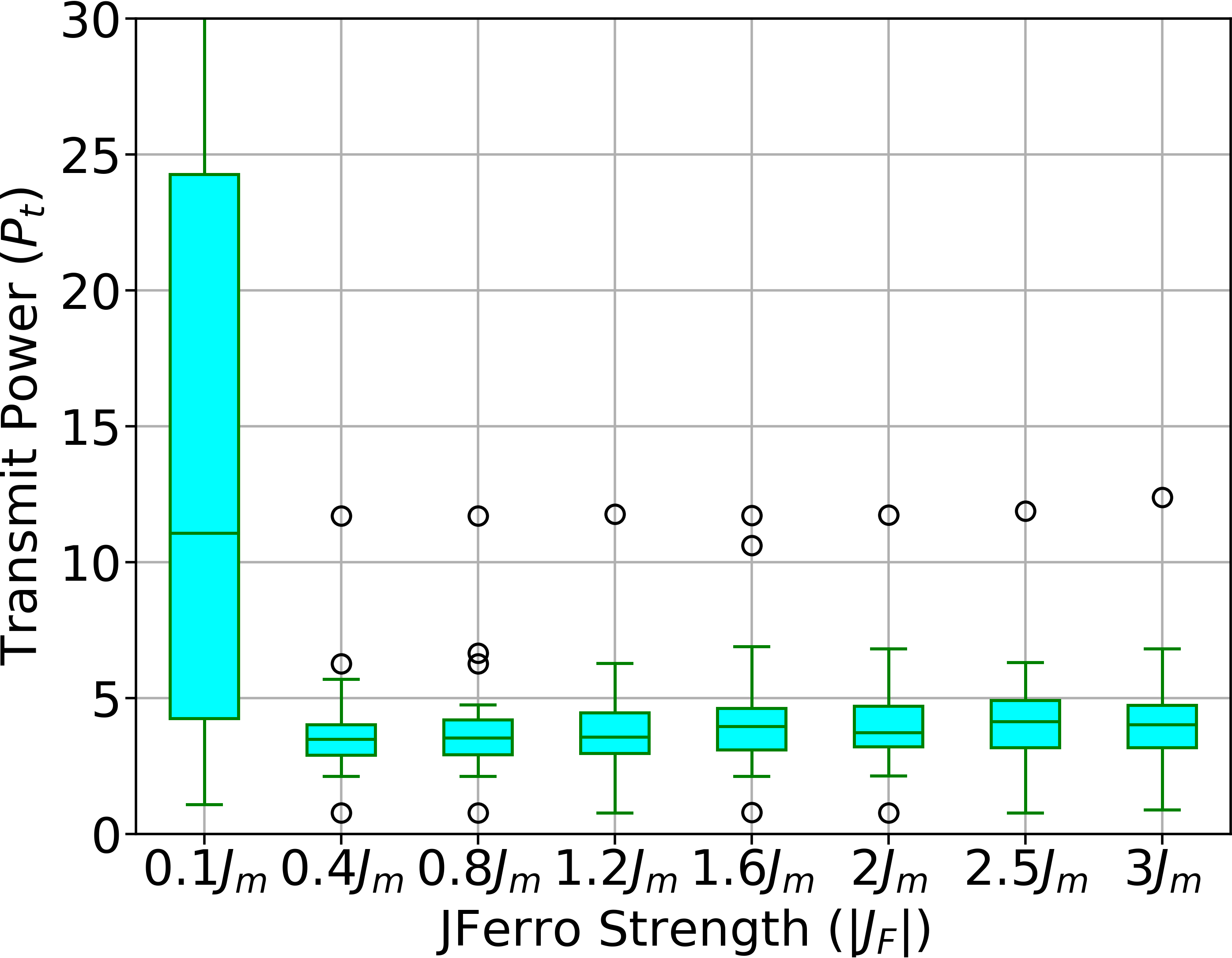}
\caption{Evaluation of the overall transmission power ($P_t$) amplitudes over various choices of $|J_F|$ at $T_a$ = $100$~$\mu$s and 10~dB channel SNR. The plot shows that QAVP's performance variance is not significant for $J_F$ values above $0.4J_m$.}
\label{f:poweranalysis}
\end{figure}

We evaluate QAVP's performance over an $10\times 8$ MIMO system with 20 QAVP problem instances. Our channel matrix \textbf{H} is random Rayleigh fading, modulation scheme is BPSK, and wireless channel noise is Gaussian. Let $J_m$ be the maximum coupler strength value of our QUBO problem at hand. 

In Fig.~\ref{f:poweranalysis} we investigate the sensitivity of chain strengths $\abs{J_F}$ over the overall transmission power $P_t$. For this evaluation, we set a high annealing time of $100~\mu s$ and a moderate SNR of $10~dB$ to ensure minimal disturbance from the time limit and the channel SNR respectively. We observe that the performance of QAVP varies with $J_F$ and that the sensitivity of QAVP performance beyond $|J_F|$ = $0.4J_m$ is not significant. The average $P_t$ is much higher for $|J_F| < 0.4J_m$, and the variance in performance of QAVP is barely distinguishable for higher values of $|J_F| > 0.4J_m$. While the optimal value of $J_F$ depends on the parameters of the system (\textit{e.g.,} modulation and MIMO size), we empirically determine that $|J_F| = 1.2J_m$ is the optimal setting for $6\times 6$ and $12\times 12$ MIMO systems with 64 QAM modulation scheme.
\begin{figure}
    \centering
    \includegraphics[height=0.48\linewidth,width=\linewidth]{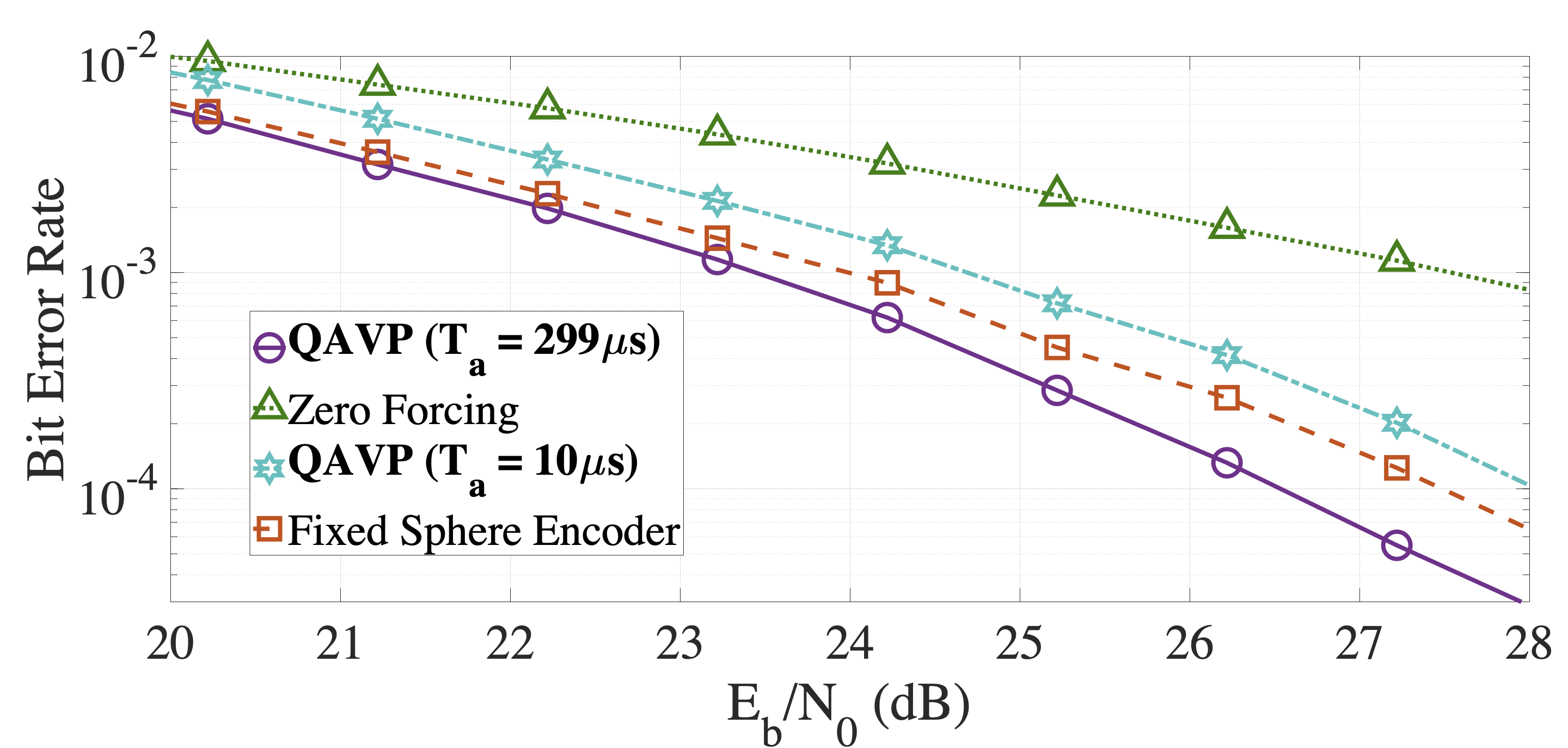}
    \caption{Downlink multi-user $6\times 6$ MIMO: BER for various QAVP anneal times $T_{a}$. We compare error performance of QAVP for various values of anneal time ($T_a$) while keeping the number of annneals ($N_a = 10^4$) fixed. We set $|J_F| = 1.2J_m$.}
    \label{fig:mimo6x6Ta}
\end{figure}


\begin{figure}
    \centering
    \includegraphics[height=0.46\linewidth,width=\linewidth]{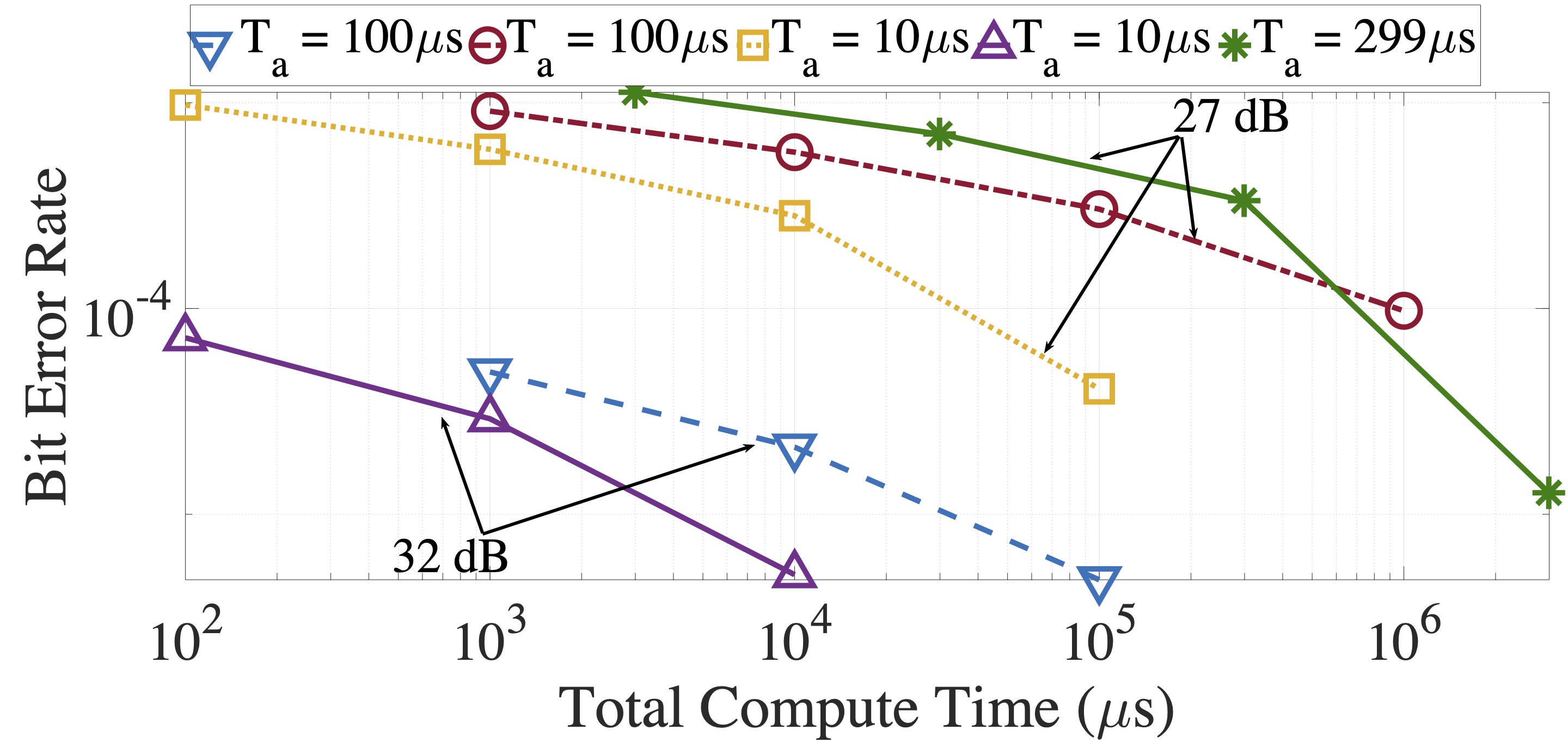}
    \caption{Downlink multi-user $6\times 6$ MIMO: Bit Error Rate vs Total Computation Time ($N_a \times T_a$). We demonstrate the BER performance, at two values of SNR, over different combinations of anneal time ($T_a$) and number of anneals ($N_a$). We set $|J_F| = 1.2J_m$.}
    \label{fig:mimo6x6TTb}
    \vspace{-10pt}
\end{figure}

\subsection{Effect of Anneal Time ($T_a$)}
Another critical QA parameter that affects the QAVP performance is the anneal time ($T_a$), along with its associated number of anneals ($N_a$). We heretofore quantify the QA performance with the total compute time: $T_a\times N_a$. Fig.~\ref{fig:mimo6x6Ta} compares the BER performance of FSE, ZF and QAVP, and Fig.~\ref{fig:mimo6x6TTb} reports the BER performance of QAVP at different QA compute times, for a $N_{t} = 6$, $N_{r} = 6$ MIMO system with 64 QAM modulation, for various choices of $T_a$ and $N_a$.


We first observe in Fig.~\ref{fig:mimo6x6Ta} that a higher anneal time leads to better BER performance from QA. Although a higher anneal time allows QA to return better quality solutions, it increases the communication latency (due to a higher computation time) as Fig.~\ref{fig:mimo6x6TTb} depicts. The tradeoff between BER performance and anneal time represents a fundamental design problem in QAVP. We observe from Figs.~\ref{fig:mimo6x6Ta} and \ref{fig:mimo6x6TTb} that QAVP achieves an acceptable BER with a low anneal time of $T_a = 10\mu s$, while keeping the computation time lower by an order of magnitude.


\subsection{End-to-end BER performance}

\begin{figure}
    \centering
    \includegraphics[height = 0.48\linewidth,width=1\linewidth]{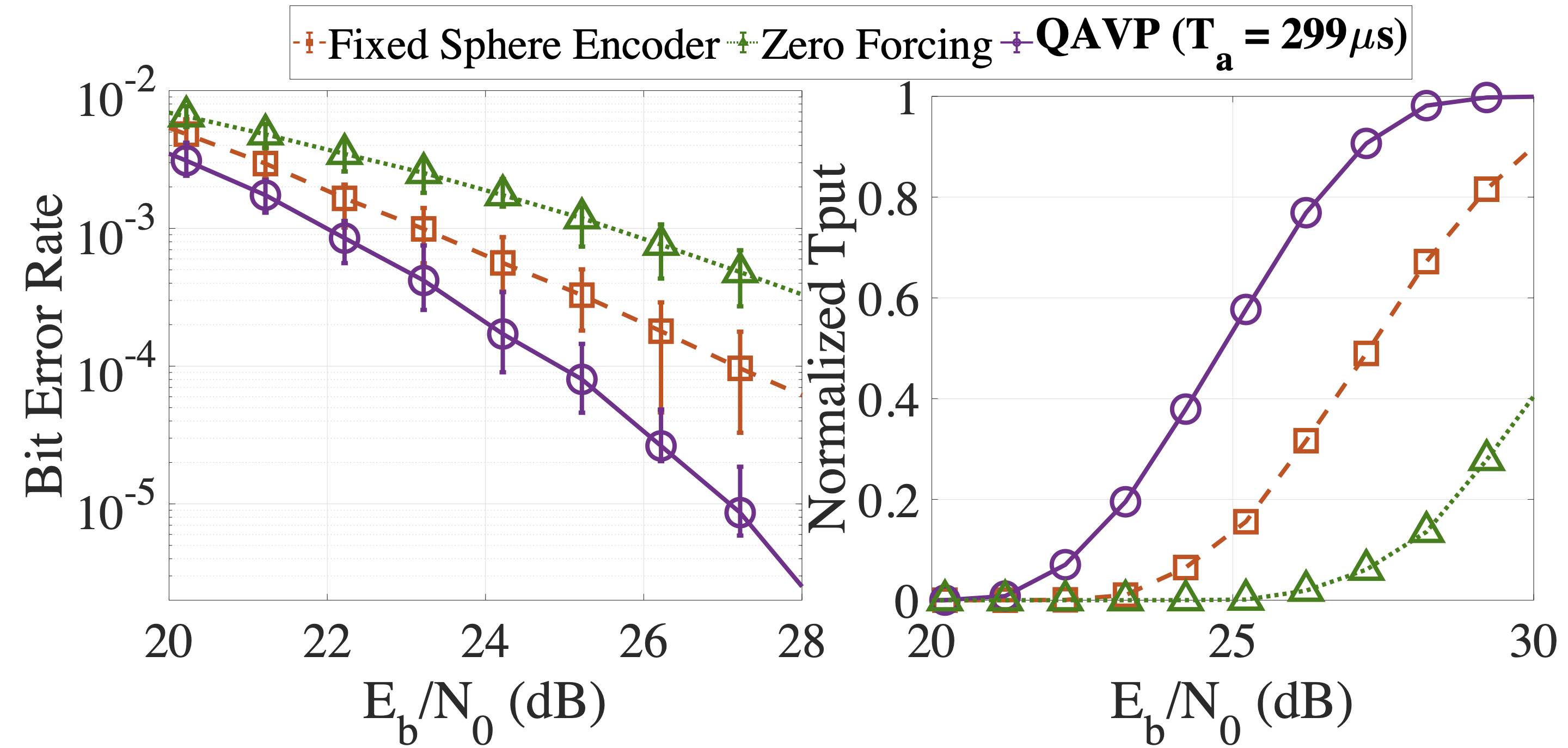}
    \caption{Downlink multi-user $12\times 12$ MIMO: BER and throughput performance. We compare the BER performance of QAVP against FSE and ZF. We set $|J_F| = 1.2J_m$, $T_a = 299\mu$s and $N_a = 10^4$.}
    \label{fig:mimo12x12}
    \vspace{-10pt}
\end{figure}

In this section, we consider a large $12\times 12$ MIMO system with
Rayleigh fading channel, White Gaussian noise, and 64 QAM modulation. Fig.~\ref{fig:mimo12x12} reports results (with $|J_F| = 1.2J_m$, $T_a = 299$~$\mu$s, $N_a = 10^4$) for mean BER and normalized throughput (ratio of achieved to maximum possible throughput), with packets size of 1500 bytes, for uncoded data transmission. We observe from Fig.~\ref{fig:mimo12x12} that QAVP outperforms ZF and FSE by achieving a 1--2 orders of magnitude lower BER at $E_b/N_0$ of 27dB. We see from the throughput curves that QAVP achieves a 3~dB gain over FSE and around 6~dB gain over ZF. It is to note that a 12$\times$12 MIMO system with VPP, using theoretically optimal Sphere Encoder, is practically infeasible due to its high computational complexity.

\section{Conclusion}
\label{sec:concl}
In this paper we propose QAVP, a novel technique that performs Vector Perturbation Precoding using Quantum Annealing. We evaluate QAVP on a real DW2Q QA device over a variety of machine parameters. Our studies show that for large MIMO systems, if the analysis is restricted to computation time, QAVP can outperform practically feasible state-of-the-art techniques like FSE and ZF algorithms for VPP. While prior work investigates QA technology for problems in the uplink \cite{10.1145/3372224.3419207, qaMIMO1}, our studies investigate its potential in the downlink. Our analysis disregards the engineering and system integration overheads of currently available commercial QA systems (e.g. latency, programming time and thermalization times between the runs), since they can be optimized heavily if the system is built for a specific application deployment. Nevertheless, the techniques we propose in this work, in the more distant future, may enable the applicability of VPP to large MIMO systems by exploiting stronger QA devices and parallelism.

\subsection{Future Work}

Our work has several possible improvements along the lines of QUBO pre-processing, problem embedding, and QA machine parameter selection. It is of interest to further refine our pre-processing step to optimize the QUBO representation based on the embedding of the problem. Further, a more sophisticated tuning and selection of QA parameters such as chain strengths and annealing times can potentially improve the QAVP performance. Porting our model to other alternative emerging Ising-model related technologies such as quantum-inspired algorithms, reverse quantum annealing, and quantum-classical hybrid approaches is also a potential work direction.


\section*{Acknowledgements}
This research is supported by National Science Foundation (NSF) Awards CNS-1824357 and CNS-1824470. Support from USRA Cycle 3 Research Oppurtunity Program allowed machine time on a D-Wave Quantum Annealer machine hosted at NASA Ames Research Center.
\bibliographystyle{ieeetr}
\bibliography{reference}
\end{document}